\documentstyle[12pt,epsfig]{article}
\begin{document}
\parskip=0pt
\parindent=125mm
DESY 98-037\par
HIP-1998-07/TH\par
TECHNION-PH-11\par
\bigskip
\bigskip
\begin{center}
\parskip=20pt{\huge Searching for new physics in $b\to s s \bar d$ decays}
\bigskip
\bigskip
\bigskip
\bigskip
\bigskip

{\large Katri Huitu$^a$,  Cai-Dian L\"u$^b$\footnote{Alexander von Humboldt 
research fellow.}, 
Paul Singer$^c$ and Da-Xin Zhang$^a$ }

$^a$ 
Helsinki Institute of Physics, P.O.Box 9,
FIN-00014 University of Helsinki, Finland
\\
$^b$ II. Institut f\"ur Theoretische Physik, Universit\"at Hamburg,
22761 Hamburg, Germany
\\
$^c$ 
Department of Physics, Technion- Israel Institute of Technology,
Haifa 32000, Israel

\end{center}

\parindent=33pt
\bigskip
\begin{abstract}
For any new physics possibly emerging in the future B experiments,
the problem is how to extract the signals 
from the SM background.
We consider the decay $b\to s s\bar d$
which is very small in the SM.
In the MSSM this decay is possibly  accessible
in the future experiments.
In the supersymmetric models with R-parity violating couplings,
this channel is not strictly constrained,
thus being useful in obtaining bounds on the lepton-number
violating couplings.
A typical candidate for the suggested search is the 
$B^-\to K^-K^-\pi^+$ mode.
\end{abstract}

 12.15.Ji, 13.25.Hw, 12.60.Jv
\newpage
\parskip=5pt
Rare $b$ decays offer a good opportunity to discover
new physics beyond the standard model (SM).
Many investigations have been done in the past years
on the predictions of processes induced by 
flavor changing neutral current (FCNC) interactions,
both within the SM 
 and beyond \cite{fcnc-bey}.
One of these FCNC induced processes, 
$b\to s\gamma$, 
has been measured \cite{bsr-exp}
and the branching ratio is comparable with
the SM prediction \cite{bsr-sm} which however still 
contains significant uncertainties.
Thus it is hard to make any definite conclusion of signals of new physics.
This is also true in most of the channels like
$b\to s q\bar q$ \cite{bsqq} and $b\to s l\bar l$ \cite{bsee},
due to the theoretical uncertainties.

In a recent study,
Gabbiani {\it et al.} \cite{sbox} considered the non-leptonic processes 
$s\to d q\bar q$ ($q=u, d, s$) more completely in the
supersymmetric model by including also the box diagrams,
in addition to the penguin contribution calculated before.
Bounds from $b\to s\gamma$,
$B\bar B$ and $K\bar K$ mixings are considered
and their conclusion is that the box diagrams cannot be neglected
in the non-leptonic transitions.

Here we will consider a novel channel $b\to ss\bar d$
which turns out to be exceedingly small in the SM.
In the SM,
this process can be induced by box diagrams with
the up-type quarks and weak bosons inside the loop.
Due to the strong GIM-suppression and the small CKM angles
involved, 
the $W$-box contribution is found to be very small.
We perform a simple estimate and find that within
the SM,
\begin{eqnarray}
\Gamma=\frac{m_b^5}{48(2\pi)^3}
\left|\frac{G_F^2}{2\pi^2}m_W^2
V_{tb}V_{ts}^* \left [V_{td}V_{ts}^*f\left(\frac{m_W^2}{m_t^2}\right)
+V_{cd}V_{cs}^* \frac{m_c^2}{m_W^2}
g\left(\frac{m_W^2}{m_t^2},\frac{m_c^2}{m_W^2}\right)\right]
\right|^2,
\end{eqnarray}
where
\begin{eqnarray}
f(x)&=&\frac{1-11x+4x^2}{4x(1-x)^2}
-\frac{3}{2(1-x)^3}{\rm ln}x,\\
~g(x,y)&=& \frac{4x-1}{4(1-x)} +\frac{8x-4x^2-1}{4(1-x)^2}\ln x -\ln y .
\nonumber
\end{eqnarray}
In eqn. (1) the  $\displaystyle\frac{m_c^2}{m_W^2}$ term is
numerically about one half of the highly CKM-suppressed contribution
at the amplitude's level.
We have dropped in (1) the kinematics 
dependent contribution
which is smaller than $10\%$ of the term proportional to
$\displaystyle\frac{m_c^2}{m_W^2}$.
Even though the relative phase between the two contributions
in (1) is unknown,
the branching ratio is always less than $10^{-11}$,
far beyond the designed ability of B-factories.
By comparing with the analogous processes $B^0\bar B^0$ and
$K^0 \bar K^0$ mixings \cite{box},
we suppose that
including QCD correction will not change the
value greatly.
Furthermore,
the so-called ``dipenguin'' \cite{dipeng}
is only part of the ${\cal O}(\alpha_s)$
corrections to the lowest order $W$-box diagram,
and is thus less important.
In order to consider new physics,
this is a clean and useful channel.
If this process were observed at future experiments,
we would be confident that there is new physics involved.

In the minimal supersymmetric standard model (MSSM),
this transition can be induced by the squark-gaugino (or higgsino) box 
diagrams.
Since $\cot \beta$ is constrained to be small
in the MSSM,
there is no large contribution from the charged Higgs box diagrams
and we will not consider it further.
An alternative mechanism for this channel 
in the supersymmetric models
is through the $R$-parity violating couplings.
These two seem to be the only ones capable in
mediating this decay within the  supersymmetric models without
strong suppression.
The non-supersymmetric models like the two Higgs doublet models,
including the so-called Model-III \cite{2hd3},
are worth a separate investigation.

To simplify our discussion,
we consider only the squark-gluino box
which is generally the dominant contribution.
Following the mass-insertion approximation \cite{mass-insert,sbox},
 universal squark masses are assumed,
and the squark mixings
are described by  
the off-diagonal elements in 
the mass squared  matrices.
We keep only the left-handed sector
in the squark mixing,
following the observation made in \cite{bsr-susy}
that the left-right and the right-right
sectors are more strongly constrained.
The effective Hamiltonian is then
\begin{eqnarray}
{\cal H}=-\frac{\alpha_s^2\delta_{12}^{d*}\delta_{23}^{d}}{216m_{\tilde d}^2}
[24xf_6(x)+66{\tilde f}_6(x)] (\bar s \gamma^\mu d_L)(\bar s \gamma_\mu b_L),
\end{eqnarray}
where 
$x=m_{\tilde g}^2/m_{\tilde d}^2$,
 and the functions $f_6(x)$ and ${\tilde f}_6(x)$
can be found in \cite{sbox}.
$\delta_{ij}^d$ parameterizes the mixing between the down-type
left-handed squarks.
The decay width is calculated as
\begin{eqnarray}
\Gamma=\frac{\alpha_s^4|\delta_{12}^{d*}\delta_{23}^{d}|^2m_b^5}
{48(2\pi)^3 m_{\tilde d}^4 }
\left[\frac{2}{9} x f_6(x)+\frac{11}{18}{\tilde f}_6(x)\right]^2.
\end{eqnarray}
At present,
the strongest bounds on the squark mixing parameter
$\delta_{12}^d$ comes from $K\bar K$ mixing,
and $\delta_{23}^d$ from $b\to s\gamma$ \cite{sbox}.
These bounds are obtained using $\Delta m_K <3.521 \times 10^{-15}$GeV
and $BR(b\to s\gamma)<4\times 10^{-4}$. They depend on $x$.
Using these bounds,
we plot in Figure 1 the maximum branching ratio 
of  $b\to ss\bar d$ depending on $x$.
When doing numerical calculations, our parameters are chosen as
$m_{\tilde d}=500$ GeV, $\tau_B=1.59$ ps, $f_K=160$ MeV, $m_b=4.5$ GeV. 
Note that QCD corrections are less important in the MSSM \cite{sbox}. 
\begin{figure}
    \epsfig{file=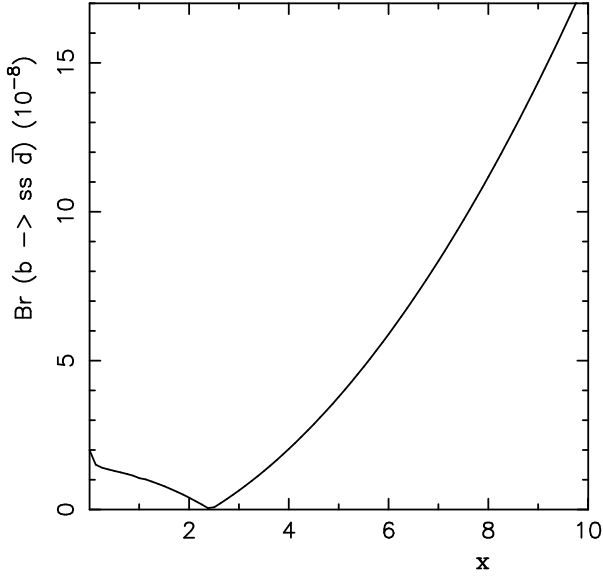,bbllx=2cm,bblly=7.5cm,bburx=21cm,bbury=19cm,%
width=12cm,angle=0}
    \caption{The branching ratio of $b\to ss\bar d$ as a function of 
$x=m_{\tilde g}^2/m_{\tilde d}^2$ in MSSM without R-parity violation,
when the squark-gluino box diagram is included.
Region above the line has been excluded by the present data
on $b\to s\gamma$ and on $\Delta m_K$.}
    \label{fig}
\end{figure}

The MSSM can be extended by including 
R-parity violating interactions.
The term in the R-parity violating part of the superpotential, which is 
relevant here is
\begin{eqnarray}
W=\lambda '_{ijk} L_iQ_jd_k,
\label{wrviol}
\end{eqnarray}
where $i,j,k$ are indices for the families and $L,Q,d$ are,
under the SM gauge group,
the superfields for the lepton doublet,
the quark doublet and the down-type quark singlet, respectively.
$\lambda '$ is a dimensionless coupling.
The transition $b\to ss\bar d$ can be induced also by the lepton number 
violating interactions in $W$.
Following the notations in \cite{choud},
the tree level effective Hamiltonian is
\begin{eqnarray}
{\cal H}=-\sum_n \frac{f_{\rm QCD}}{m_{\tilde \nu_n}^2}
(\lambda_{n32}^\prime \lambda_{n21}^{\prime *} \bar s_R b_L \bar s_L d_R
+\lambda_{n12}^\prime \lambda_{n23}^{\prime *} \bar s_R d_L \bar s_L b_R).
\label{tt}
\end{eqnarray}
The QCD corrections to the left-right operators in eqn.(\ref{tt})
 has been found to be important \cite{bagger}. 
The next-to-leading order QCD corrections are also available \cite{ciu}.
For simplicity, here we only include the leading order QCD corrections 
which are given by a scaling factor 
\begin{equation}
f_{\rm {QCD}}=\left( \frac{ \alpha_s(m_b)}{\alpha_s(m_t)} \right)^
{\frac{24}{23}}  \left( \frac{ \alpha_s(m_t)} {\alpha_s(m_{\tilde \nu_n})}
 \right)^{\frac{24}{21}}
\end{equation}
for $m_{\tilde \nu_n}>m_t$, and by
\begin{equation}
f_{\rm {QCD}}=\left( 
\frac{ \alpha_s(m_b)}{\alpha_s(m_{\tilde \nu_n})} \right)^{\frac{24}{23}} 
\end{equation}
for $m_{\tilde \nu_n}<m_t$.
Using $m_{\tilde \nu_n}=100$ GeV, we estimate $f_{\rm{QCD}}\simeq 2$.

Then the decay rate induced by the R-parity violating couplings is
\begin{eqnarray}
\Gamma =\frac{m_b^5f_{\rm {QCD}}^2}{512(2\pi)^3}
\left(\left|\sum_n \frac{1}{m_{\tilde \nu_n}^2}
\lambda_{n32}^\prime\lambda_{n21}^{\prime *}\right|^2
+
\left|\sum_n \frac{1}{m_{\tilde \nu_n}^2}
\lambda_{n12}^\prime \lambda_{n23}^{\prime *}\right|^2\right).
\end{eqnarray}
Note that the couplings are not strongly constrained by the
present experiments \cite{rp-rev}:
\begin{eqnarray}
|\lambda_{132}^\prime \lambda_{121}^{\prime *}|<0.34\times 0.035,&
|\lambda_{112}^\prime \lambda_{123}^{\prime *}|<0.02\times 0.20,\nonumber\\
|\lambda_{232}^\prime \lambda_{221}^{\prime *}|<0.36\times 0.18,&
|\lambda_{212}^\prime \lambda_{223}^{\prime *}|<0.09\times 0.18,\nonumber\\
|\lambda_{332}^\prime \lambda_{321}^{\prime *}|<0.48\times 0.20,&
|\lambda_{312}^\prime \lambda_{323}^{\prime *}|<0.10\times 0.20
\end{eqnarray}
if using $m_{\tilde \nu_n}=100$ GeV,
and we have
\begin{eqnarray}
\sum_{n} \sqrt{
|\lambda_{n32}^\prime \lambda_{n21}^{\prime *}|^2+
|\lambda_{n12}^\prime \lambda_{n23}^{\prime *}|^2} <0.1,\label{rbound}
\end{eqnarray}
which is too weak to constrain the present mode.
Thus a search for this decay mode will improve our
knowledge on these couplings.
At present,
an analysis of this transition at the level
of branching ratio $10^{-4}-10^{-5}$
is realistic,
and a negative result will improve the bound in (\ref{rbound})
to $10^{-4}$.
Note that the stricter constraints  
on $|\sum_n \frac{1}{m_{\tilde
\nu_n}^2}\lambda_{n32}^\prime\lambda_{n23}^{\prime *}|$
from $\Delta M_B$ and 
$|\sum_n \frac{1}{m_{\tilde
\nu_n}^2}\lambda_{n12}^\prime\lambda_{n21}^{\prime *}|$
from $\Delta M_K$ \cite{choud} are independent from the present combination
of the couplings.

Next we consider the experimental implications of the discussed channel.
In the MSSM,
the branching ratio of this decay is smaller than $10^{-7}-10^{-8}$,
which is difficult to  reach at the B-factories,
but hopefully is possible at Hera-B or at LHC.
In the MSSM with R-parity violation,
there is no strict constraint on the mode,
and the branching ratio might be quite large.
A search for this mode will 
help to improve the bounds on these $\lambda'$-type
R-parity violating couplings.

Typical final exclusive processes of $b\to ss\bar d$
include $B^\pm\to K^\pm K^0(\bar K^0)$,
which are difficult to separate
from the standard penguin process $b\to ds\bar s$ 
through $K^0\bar K^0$ mixing.
Although the interference of these
two sources of the final states are novel in the
study of the phenomena such as CP violation,
these channels are not suitable for the 
direct search for the new physics. 
However,
the three-body mode of the charged B decays
like $B^-\to K^- K^-\pi^+$,
either a direct three-body transition or
through a $\bar K^{0*}$-like resonance,
will be a clear signal for  this mode. 
In the neutral B decays,
the channel $\bar B^0\to K^- K^-\pi^+\pi^+$ 
is also a clear signal.
Similar consideration also applies in other
$K^\mp K^\mp +({\rm no~ strange})$ final states,
which can be searched at the B-factories.
Thus we suggest to search for the signals of multi-body
channels in the B decays,
which will be useful 
in bounding the R-parity violating couplings at present, 
and in discovering physics beyond the SM in the future.

To estimate the semi-inclusive rate of 
$B\to K^\mp K^\mp +({\rm no~ strange})$,
we assume that  the multi-body
transitions are dominated by the two-body channels
which contain the excited states of the $K$ mesons.
Because of the short lifetimes of these excited states
the mixing effects between the neutral excited states 
are totally negligible.
We denote an excited $K$ as $K^*$ and the
ground state as $K$ and estimate the possibility
of  $K^*$ decays into a charged $K$ by isospin analyses.
In the decay $K^*\to K +({\rm no~ strange})$
the isospin of the nonstrange system can be $I=1$ or $I=0$.
In the $I=1$ channels of the neutral $K^*$ decays,
the possibility for a final charged $K$ is $2/3$.
This possibility for the charged $K^*$ is $1/3$.
In the $I=0$ channels,
all the charged $K^*$'s decay into charged final $K$'s
while for the neutral $K^*$'s there is no charged $K$ in
the final states.
To avoid model calculations for the individual isospin amplitudes,
we simply average over both the charged and neutral
$K^*$'s and over all the channels $K^*\to K$,
and we expect that about half of 
the decays $K^*\to K$ have charged $K$'s in the
final states.
Thus in the $B\to K^*K^*$ decays induced by $b\to ss\bar d$,
a quarter of these transitions
materialize as $B\to K^\mp K^\mp +({\rm no~ strange})$.
Similar analysis applies for the decays  $B\to K^*K$
with a smaller  possibility
to have two charged  $K$'s in the final states;
however, 
we can expect that there are less $K^*K$ than $K^*K^*$ channels,
and the individual transition $B\to KK$ is even less dominant.
We conclude that the estimated $1/4$ possibility of 
having two charged (same sign) $K$'s roughly works,
and  the semi-inclusive process $B\to K^\mp K^\mp +({\rm no~ strange})$ 
consists about $1/4$ of all the $b\to ss\bar d$ transitions.

Finally,
let us consider a similar process $b\to dd\bar s$
due to the same mechanisms.
An interesting exclusive channel of this
process is $B^-\to K^+\pi^-\pi^-$. 
In the SM,
this process suffers even stronger suppression
than $b\to ss\bar d$ (by a factor of $|V_{td}/V_{ts}|$ in the amplitude).
In the MSSM,
the decay rate is proportional to
the more strongly constrained $|\delta_{21}^{d*}\delta_{13}^d|^2$.
Thus its upper bound is much smaller (only $10^{-4}$ of $b\to ss\bar d$).
However, 
the hope that $b\to dd\bar s$ can be induced by
the $\lambda '$-type $R$-parity violating couplings
is still alive.
By comparing the involved $\lambda '$ combination with
that in $b\to ss\bar d$,
again,
there exists only a very loose bound
\begin{eqnarray}
\sum_{n}\sqrt{
|\lambda_{n31}^\prime \lambda_{n12}^{\prime *}|^2+
|\lambda_{n21}^\prime \lambda_{n13}^{\prime *}|^2} <0.05,\label{rbound2}
\end{eqnarray}
and the available data can be used to improve this bound
to the order of $10^{-4}$.

We thank Kenneth \"Osterberg for helpful discussions.
The work of KH and DXZ is partially
supported by the Academy of Finland (no. 37599).
The work of PS is partially supported by the Fund for
Promotion of Research at the Technion.

\end{document}